\documentclass[aps,twocolumn,superscriptaddress,prl,10pt]{revtex4-1}

\usepackage{amsmath,amsthm,amssymb,bm,hyperref,graphicx,color,float}
\allowdisplaybreaks

\newcommand{\be}{\begin{equation}}
\newcommand{\ee}{\end{equation}}

\newcommand{\inti}{\int_{-\infty}^{+\infty}}

\newcommand{\6}{\partial}

\begin{document}

\title{Universality and quantum criticality of the one-dimensional spinor Bose gas}

\author{Ovidiu I. P\^{a}\c{t}u}
\affiliation{Institute for Space Sciences, Bucharest-M\u{a}gurele, R 077125, Romania}
\author{Andreas Kl\"umper}
\affiliation{Fakult\"at f\"ur Mathematik und Naturwissenschaften, Bergische
 Universit\"at Wuppertal, 42097 Wuppertal, Germany}
\author{Angela Foerster}
\affiliation{Instituto  de  F\'{i}sica  da  UFRGS,  Av.  Bento Gon\c{c}alves  9500,  Porto  Alegre,  RS,  Brazil}


\begin{abstract}

We investigate the universal thermodynamics of the two-component one-dimensional
Bose gas with contact interactions in the vicinity of the quantum critical point
separating the vacuum and the ferromagnetic liquid regime. We find that the quantum
critical region belongs to the universality class of the spin-degenerate
impenetrable particle gas  which, surprisingly, is very different from
the single-component case and identify its boundaries with the peaks of the specific heat.
In addition, we show that the compressibility Wilson ratio, which quantifies the
relative strength of thermal and quantum fluctuations, serves as a good discriminator
of the quantum regimes near the quantum critical point. Remarkably, in the
Tonks-Girardeau regime the universal contact develops a pronounced minimum,
reflected in a counterintuitive narrowing of the momentum distribution as we
increase the temperature.  This momentum reconstruction, also present at low and
intermediate momenta, signals the transition from the ferromagnetic to the
spin-incoherent Luttinger liquid phase and can be detected in current
experiments with ultracold atomic gases in optical lattices.

\end{abstract}

\maketitle

Experiments with ultracold atomic gases allow for the creation and
manipulation of various physical systems with an unprecedented degree of
control over interaction strength, statistics of the constituent particles and
dimensionality \cite{BDZ}. This control materialized in the experimental
realization of many low-dimensional systems whose physics is well captured by
many-body integrable models allowing for a parameter free comparison between
theory and experiment \cite{CCGOR,GBL}. The need for accurate theoretical
predictions is particularly stringent in low-dimensions due to the enhancement
of quantum fluctuations which invalidates our physical intuition based on the
free-particle picture and mean-field theory. In the case of integrable models
powerful techniques associated with Bethe Ansatz \cite{KBI,EFGKK} can be employed
to analytically investigate the physical properties of such systems for all values
of the interaction strength including the strong coupling limit.

One-dimensional (1D) multi-component systems in which the constituent
particles have a variable number of internal ``pseudospin" states present very
rich physics exhibiting quantum regimes not found in higher dimensions or in
the case of their single component counterparts
\cite{CCGOR,GBL,WWD,Pagano}. As is well known, the single
  component Bose gas in the Tonks-Girardeau regime with infinitely strong
  coupling realizes an impenetrable particle gas which is fully
  described by spinless free fermions. We will see that the spinor gases in
  the very strong coupling regime behave very differently from free fermions.

At zero temperature these systems present Quantum Phase Transitions (QPT)
which are driven by varying certain nonthermal parameters (chemical potential,
magnetic field, pressure, etc.). In the vicinity of the Quantum Critical Point
(QCP) quantum and thermal fluctuations couple strongly characterizing
the Quantum Critical (QC) regime \cite{S}.
In the phase space the QC region has a distinct V-shape fanning out from the QCP
\cite{S,CS} at low-temperatures and in this region the thermodynamics of the
system is universal and scale invariant.

Motivated by the recent experimental confirmation \cite{Guan1} of quantum
criticality in the Lieb-Liniger model \cite{LL} in this Letter we theoretically
investigate the more complex problem posed by the two-component generalization
known as the spinor Bose gas \cite{Y1,G1,LGYE}.
We provide an analytical description of the
universal thermodynamics and determine the location of the QCP, the critical
exponents and the boundaries of the critical region. The
  analytical description is derived for arbitrary number of spin
  components. It is universally valid for bosons and fermions. It also
  explains the first order transition in the finite density regime with jump
  of the magnetization at zero magnetic field.  In addition, for the
  two-component Bose case we perform a detailed analysis of the universal
  contact \cite{Tan1,OD,BZ,CJGZ} and show that in the Tonks-Girardeau regime
  it presents a pronounced minimum signalling the transition from the
  ferromagnetic \cite{ZCG1,AT,MF,KG,ZCG2} to the spin-incoherent liquid phase
  \cite{CZ1,FB,F}. This transition is accompanied by a significant momentum
  reconstruction which can be experimentally detected.

We consider a system of 1D bosons with two internal ``pseudospin" states
denoted by $\uparrow$ and $\downarrow$ with spin-independent contact
interactions. In second quantization the Hamiltonian is $H=\int
\mathcal{H}(x)\, dx$ with density
\begin{align}\label{ham}
\mathcal{H}(x)&=\sum_{\sigma=\{\uparrow,\downarrow\}} \left(\frac{\hbar^2}{2m}
\6_x\Psi^\dagger_\sigma(x)\6_x\Psi_\sigma(x)
-\mu_\sigma\Psi^\dagger_\sigma(x)
\Psi_\sigma(x)\right)\,\nonumber\\
&\ \ \ \ \  +\frac{\hbar^2 c}{2m} \sum_{\sigma,\sigma'=\{\uparrow,\downarrow\}} \Psi^\dagger_\sigma(x)
\Psi^\dagger_{\sigma'}(x)\Psi_{\sigma'}(x)\Psi_\sigma(x)
\end{align}
where $\Psi_\sigma(x), \Psi_\sigma^\dagger(x)$ are bosonic fields
satisfying
$[\Psi_\sigma(x),\Psi^\dagger_{\sigma'}(y)]=\delta_{\sigma,\sigma'}\delta(x-y)$,
$m$ is the mass of the particles, $\mu_\sigma$ are chemical potentials and the
coupling constant can be expressed in terms of the 1D scattering length via
$c=-2/a_{1D}$ \cite{Ol}. In the following it will be useful to introduce
$\mu=(\mu_\uparrow+ \mu_\downarrow)/2$ and  $H=(\mu_\uparrow-\mu_\downarrow)/2$.
In units of $\hbar=2m=1$ the
dimensionless coupling constant is $\gamma=c/n$ with $n$ the total density and
the Fermi temperature is $T_F=\pi^2 n^2$. The system is weakly (strongly) interacting
when $\gamma\ll 1\,  (\gamma\gg1)$.

The Hamiltonian (\ref{ham}) is integrable \cite{Y1,G1} and the ground state
and low-lying excitations were derived using the Nested Bethe Ansatz
\cite{LGYE}. Compared with the fermionic model, for which the ground-state is
antiferromagnetic, in this case the ground-state is fully polarized
(ferromagnetic) which is a general characteristic of bosonic 1D models with
spin-independent interactions \cite{EL}. At low temperatures the longitudinal
low-lying excitations are plasmons ($\epsilon(p) \sim |p|$) but the softest
transversal low-lying excitation is a magnon with quadratic dispersion
$\lim_{p\rightarrow 0}\epsilon(p)\sim p^2/2m_*\, $ which prevents the use of
the Luttinger liquid (LL) theory ($ m_*=3 m\, \gamma/2\pi^2\, $ for large
$\gamma$ \cite{FGKS}).

Computing the thermodynamic properties of the spinor Bose gas is a notoriously
difficult task despite the integrability of the Hamiltonian (\ref{ham}). In
contrast with the single component case (the Lieb-Liniger model) the
Thermodynamic Bethe Ansatz (TBA) \cite{YY,T} description of the
multi-component model \cite{GLYZ} involves an infinite number of nonlinear
integral equations (NLIEs). Even though numerical schemes can be developed to
approximately solve such systems \cite{CKB,KC,GBT} they are rather cumbersome
and require truncations in the number of equations which can introduce
uncontrollable errors.  In \cite{KP1,PK1} two of the authors
derived an efficient thermodynamic description which made use of the lattice
embedding of the spinor Bose gas in the $q=3$ Perk-Schultz spin chain and the
Quantum Transfer Matrix \cite{EFGKK,K1,K2}. In this description the
grand-canonical potential per length is given by
$
\phi(\mu,H,T)=-\frac{T}{2\pi}\inti \ln [1+a_1(k)]+\ln [1+a_2(k)]\, dk\, ,
$
with $a_i(k)$ auxiliary functions satisfying the following system of NLIEs:
\begin{subequations}\label{aux}
\begin{align}
\ln a_1(k)&=-(k^2-\mu-H)/T\nonumber\\
&\ \   +[K_0*\ln A_1](k)+[K_2*\ln A_2](k-i\epsilon)\, ,\label{a}\\
\ln a_2(k)&=-(k^2-\mu+H)/T\nonumber\\
&\ \  +[K_1*\ln A_1](k+i\epsilon)+[K_0*\ln A_2](k)\, ,
\end{align}
\end{subequations}
where  $ A_i(k)=1+a_i(k)\, ,$ $K_0(k)=2c/(k^2+c^2)\, ,$
$K_1(k)=c/[k(k+ic)]\, ,K_2(k)=c/[k(k-ic)]$ and $[f*g](k)=\frac{1}{2\pi}\inti
f(k-k')g(k')\, dk'.$
We stress  that Eqs.~(\ref{aux}) are valid for all values of $\mu\,
,H\, ,T$ and $c$ and they can be easily  and efficiently numerically implemented
allowing for the exploration of certain regions of the relevant parameters  space
(like $H\ll T,\mu,c$)  which are inaccessible by TBA. These equations constitute
the basis of our investigations.

At $T=0$ and fixed $H$ the spinor Bose gas presents a QPT between
the vacuum and ferromagnetic liquid phase when the chemical potential reaches
the QCP.  In the vicinity of the QCP the thermodynamics of the system is
universal and scale invariant and can be described by \cite{ZH,FWGF}
\be\label{scaling} p(\mu,H,T)\sim
p_r(\mu,H,T)+T^{\frac{d}{z}+1}\mathcal{P}_H\left
(\frac{\mu-\mu_c(H)}{T^{\frac{1}{\nu z}}}\right) \ee
where $p=-\phi$ is the
pressure, $p_r$ the regular part, $z$ the dynamical critical exponent, $\nu$
the correlation length exponent, $d=1$ the dimension and $\mathcal{P}_H$ is a
universal function describing the singular part of the pressure.  The determination of the universality class (the values
of $z$ and $\nu$) and of the universal function is accomplished by performing
the following steps \cite{ZH}: i) Choose $z$ and $\nu$ ii) For these values of
$z$ and $\nu$ plot for several values of temperature the ``scaled pressure''
$(p- p_r)T^{-\frac{d}{z}-1}$ (at low-$T$ $p_r$ can be replaced by a constant
which in our case is zero). If the critical exponents are chosen correctly
then all plots should intersect at $\mu_c(H)$ identifying the QCP. iii) With
$z$, $\nu$ and $\mu_c(H)$ determined the plots of the ``scaled pressures" at
all temperatures as functions of $(\mu-\mu_c(H))/T^{\frac{1}{\nu z}}$ will
collapse to a single curve which is the universal function $\mathcal{P}_H$.

\begin{figure}
\includegraphics[width=\linewidth]{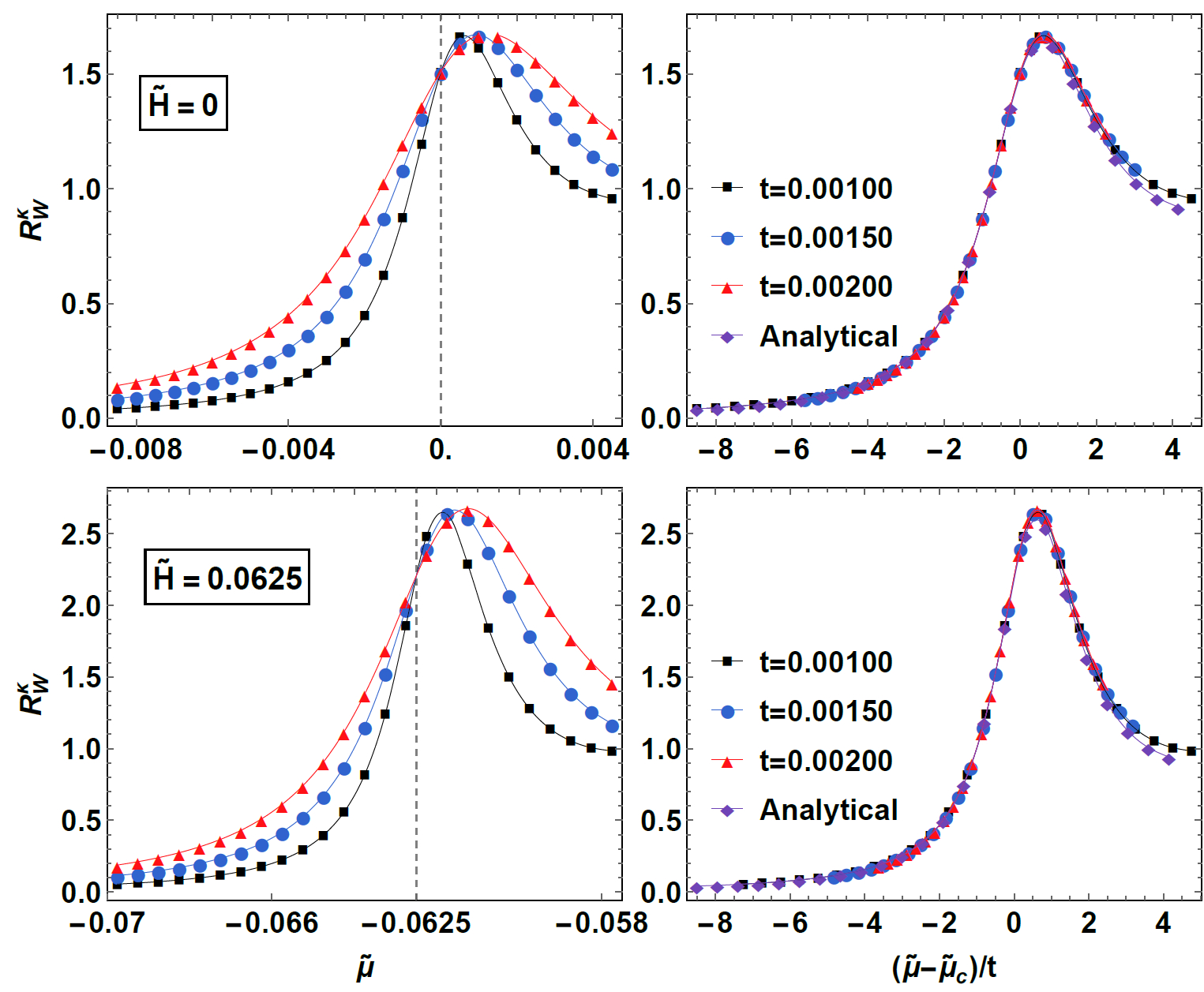}
\caption{Scaling behaviour of the compressibility Wilson ratio $R_W^\kappa$
  for $\tilde{H}=0$ (upper panels) and $\tilde{H}=0.0625$ (lower panels)
  ($\tilde H=H/c^2, \tilde\mu=\mu/c^2, t=T/c^2$ with $c=2$). The curves at
  different temperatures intersect at the QCP
  $\tilde{\mu}_c=-|\tilde{H}|$. Plots of the ratio as function of
  $(\tilde{\mu}-\tilde{\mu}_c)/t$ reveal the universal function
  $\mathcal{Q}_H(x)$.  In the right panels the magenta diamonds represent the
  analytical predictions given by Eq.~(\ref{Tak}) with $e^{-|y|}=0$ for $H\ne
  0$ and $e^{-|y|}=1$ for $H=0$.  }
\label{ScalingWilson}
\end{figure}
\begin{figure}
\includegraphics[width=\linewidth]{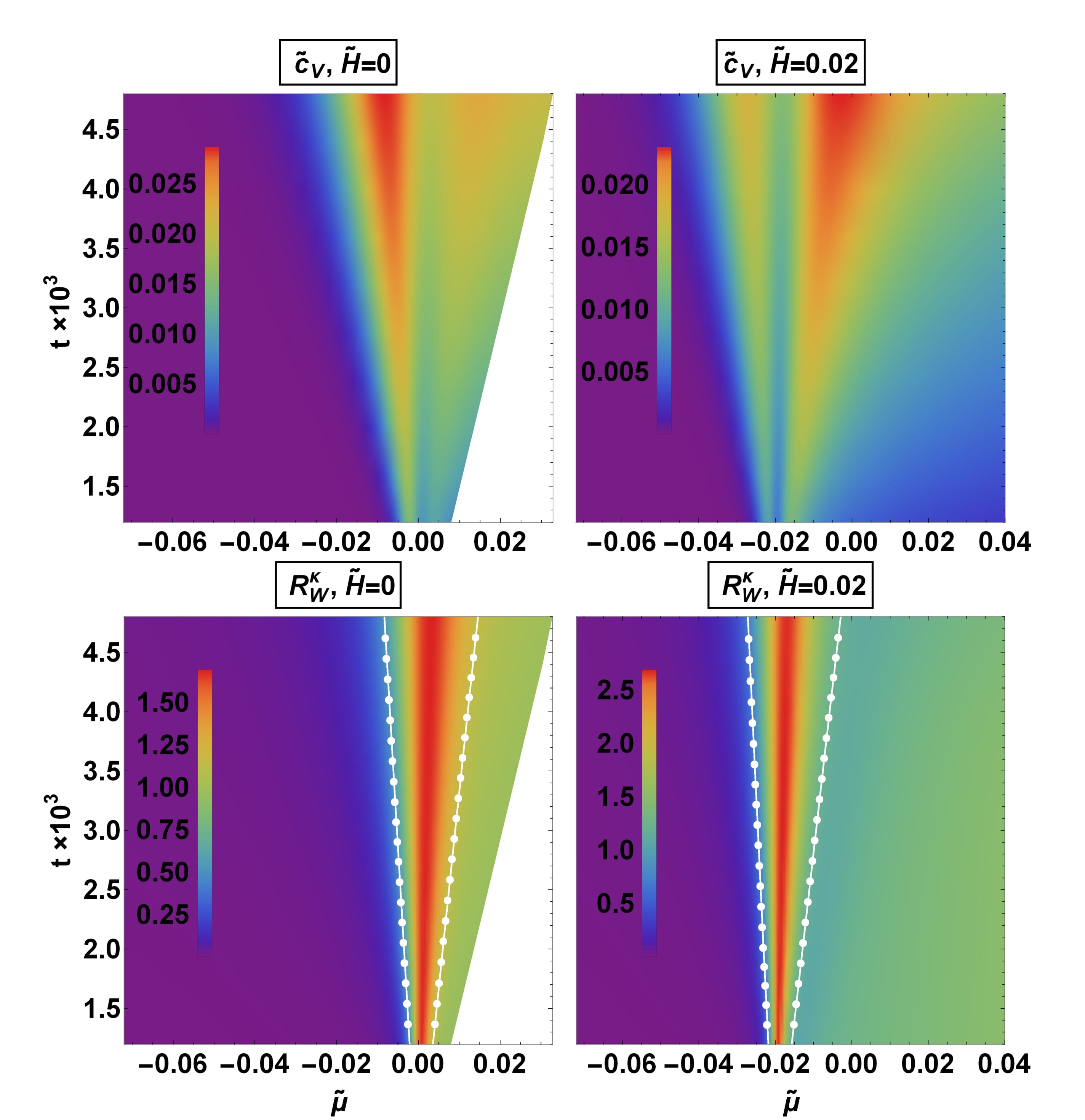}
\caption{Density plots of the grand-canonical specific heat $\tilde c_V=c_V/c$
  (upper panels) and Wilson ratio (lower panels) for $\tilde H=0$ and $\tilde
  H=0.02$ ($c=5$).  The
  specific heat presents maxima fanning out from the critical value of
  chemical potential $\mu_c=-|H|$ which determine the boundaries of the
  QC region. The white lines in the Wilson ratio plots are the
  boundaries of the QC region given by the maxima of the
  specific heat. To the left of the boundary lies the vacuum phase at finite
  temperature which can be well approximated by a classical gas and to the
  right the ferromagnetic liquid regime.
  }
\label{DensityPlots}
\end{figure}
\begin{figure}
\includegraphics[width=\linewidth]{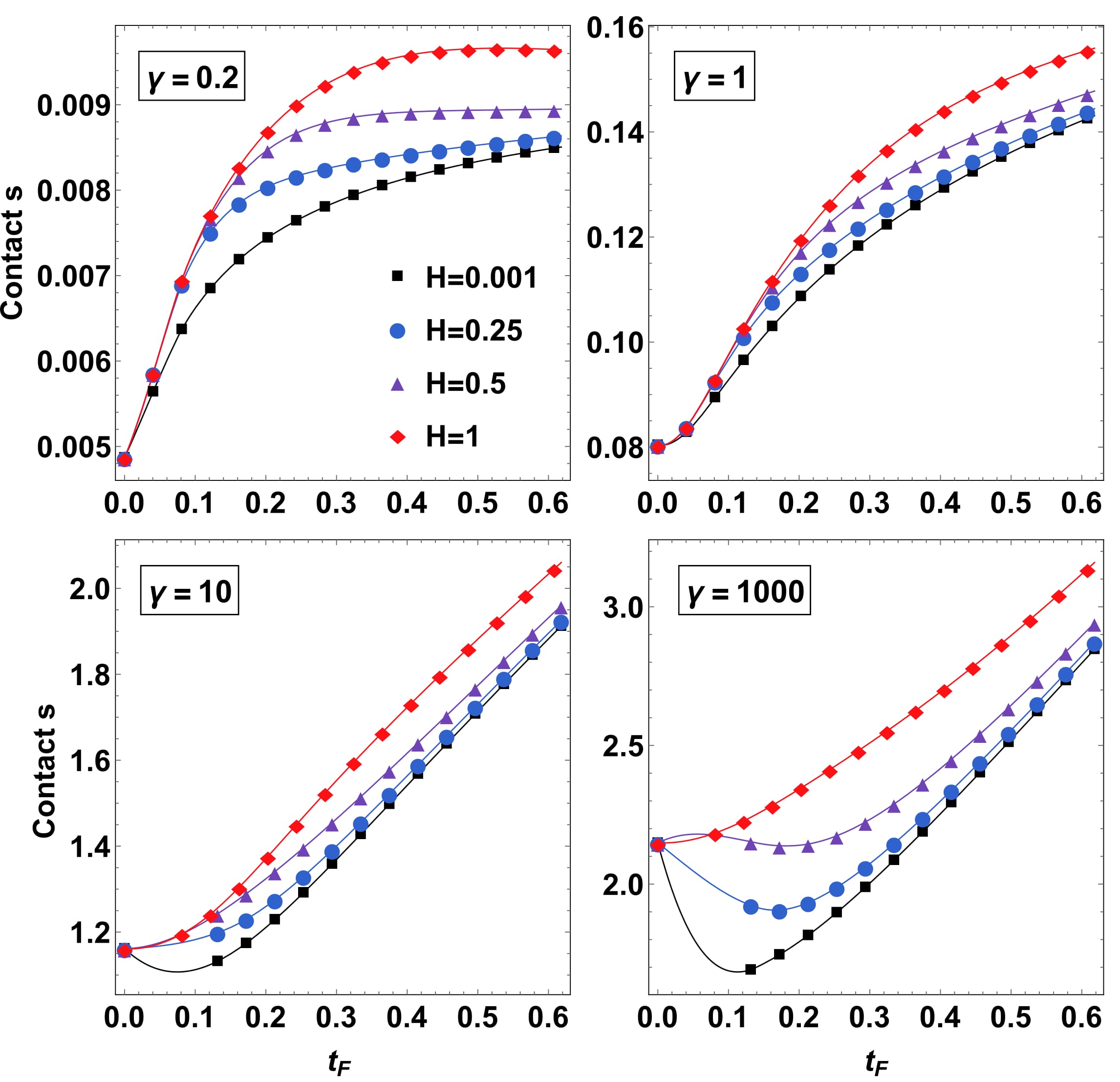}
\caption{Dependence of the dimensionless total contact $s$ on the temperature
($t_F=T/T_F\, ,\gamma=c/n\, , n=1/2$) for $\gamma=\{0.2, 1, 10, 1000\}$ and several
  values of $H$. At strong coupling and low $H$ the
  contact develops a minimum. }
\label{TanT}
\end{figure}

  Our investigations reveal \cite{SuplM} lines of QCPs at
  $\mu_c(H)=-|H|$ with critical exponents $z=2$ and $\nu=1/2$. We identify the
  universality class as the spin-degenerate impenetrable particle gas. This
  can be understood intuitively by an elementary, yet quantitative
  derivation. For any repulsive interaction, at low densities the system is in
  the strong coupling regime with finite energy states realized by wave
  functions $\psi(x_1,...,x_N,s_1,...,s_N)$ that vanish if the
  spatial coordinates of any two particles coincide. We construct such wave
  functions by first considering the fundamental regime with order of spatial
  coordinates $x_1<x_2<\cdots<x_N$. Here, we use the product ansatz
  $\phi\cdot\chi$ where $\phi(x_1,...,x_N)$ is a
  Slater determinant of plane waves with pairwise different wave numbers
  $k_1,...,k_N$ and $\chi=|\sigma_1,...,\sigma_N\rangle$ is an arbitrary spin
  part. For all other sectors with different order of spatial coordinates we
  reorder simultaneously all spatial and spin coordinates leading to the
  fundamental regime and then invoking the symmetry of the total wave
  function.

  The energy of the product states is given by the sum of the
  kinetic energy terms ($k^2$) and the Zeeman terms ($\pm H$).  We set up the
  partition function of the grand-canonical ensemble. The space of
  configurations consists of all combinations of occupations of $k$-modes:
  empty, singly occupied with spin up or spin down.
The partition function is a product over characteristic factors
$Z=\prod_k\left[1+e^{-\beta(k^2 -\mu- H)} + e^{-\beta(k^2 -\mu+ H)}\right]$
for all $k$ values allowed by the boundary conditions.
Note that this derivation with exactly the same result also works for fermions
with infinitely strong repulsion. The generalization to arbitrary spin numbers
is obvious.

The universal thermodynamics (for both fermions and bosons) in
the vicinity of the QCP is given by Takahashi's formula \cite{T}
($x=(\mu+|H|)/T\, ,y=H/T$)
\be
\label{Tak}
p=\frac{T^{3/2}}{2\pi}\inti\ln\left[1+(1+e^{-2|y|})
e^{-k^2+x}\right]\, dk\, .
\ee
Fig.~\ref{ScalingWilson} shows the  collapse of Wilson ratio (see below) data onto
the universal function analytically derived from Eq.~(\ref{Tak}).
Also, the structures of the QC regime are fully described by Eq.~(\ref{Tak}), see Fig.~\ref{DensityPlots}.
Despite its apparent simplicity, the spin degenerate impenetrable particle gas
has a rich phase diagram and surprising properties. The
critical exponents suggest that the system might be equivalent to free fermions
with the pressure given by ($x'=\mu/T$),
$p_{FF}=\frac{T^{3/2}}{2\pi}\int\ln\left[\left(1+e^{-k^2+x'+y}\right)
  \left(1+e^{-k^2+x'-y}\right)\right]\, dk\, .$ However, free fermions have a
$\mu-H$ phase diagram with four phases (vacuum, spin-up particles, spin-down
particles and the mixture of spin-up and spin-down particles) and critical
lines $\mu=\pm H$. In contrast, the spin-degenerate impenetrable particle gas has only three
phases (vacuum, spin-up particles and spin-down particles) with critical lines
$\mu=-|H|$ and $\mu>0, H=0$. At the latter line, first order transitions take
place.
Here, simply by taking the H-derivative of Eq.~(\ref{Tak}) (see also \cite{SuplM}),
we find a  magnetization with jump at zero field $m \sim \hbox{sign}H\, (\mu+|H|)^{1/2} /\pi,$
and at $T=H=0$ also residual entropy $s \sim \mu^{1/2} \ln 2/\pi\, !$ Despite the
deceiving familiar values of the critical exponents
the physics of the spin-degenerate impenetrable  particle gas is drastically
different from that of ordinary free fermions.

It was shown in \cite{YCLRG,GYFBLL,Guan2} that
the susceptibility and compressibility Wilson ratios can serve as good
discriminators of quantum phases in attractive multi-component ultracold
gases.  In our case the relevant quantity is the compressibility Wilson ratio
defined by $R_W^\kappa=\frac{\pi^2 k_B^2}{3} T \frac{\kappa}{c_V}$ with
$\kappa=-\6^2 \phi/\6 \mu^2$ the compressibility $c_V=- T \6^2 \phi/\6 T^2$
the grand-canonical specific heat at constant volume. In the vicinity of the
QCP the Wilson ratio scales like $ R_{W}^{\kappa}\sim
\mathcal{Q}_H\left(\frac{\mu-\mu_c(H)}{T}\right)$  with
$\mathcal{Q}_H(x)= \mathcal{P}_H^{''}(x)/ \left(\frac{3}{4} \mathcal{P}_H(x)-
x \mathcal{P}'_H(x)+x^2 \mathcal{P}_H^{''}(x)\right)$ a universal function. In
the vacuum phase $R_{W}^{\kappa}$ is zero and presents a sudden enhancement in
the QC region becoming almost constant in the ferromagnetic liquid phase.
Taking into account that in the grand-canonical ensemble the compressibility
and specific heat quantify the energy and particle fluctuations via $k_B T
\kappa=\langle\delta N^2 \rangle$ and $k_B T^2 c_V=\langle\delta(E-\mu
N)^2\rangle$ it is easy to understand why $R_W^\kappa$ identifies the quantum
regimes so efficiently. In the critical regime the thermal and quantum
fluctuations couple strongly resulting in the anomalous enhancement in the
vicinity of the QCP. In the upper panels of Fig.~\ref{DensityPlots} we show
that the specific heat presents a continuous set of local maxima on both sides
of the QCP, peaks that can be identified with the boundaries of the QC
region (see also \cite{Guan1,Guan2,MHO}). Depicting these boundaries in
the density plots of the Wilson ratio (lower panels of
Fig.~\ref{DensityPlots}) we see that they clearly delimitate the three quantum
regimes.

\begin{figure}
\includegraphics[width=\linewidth]{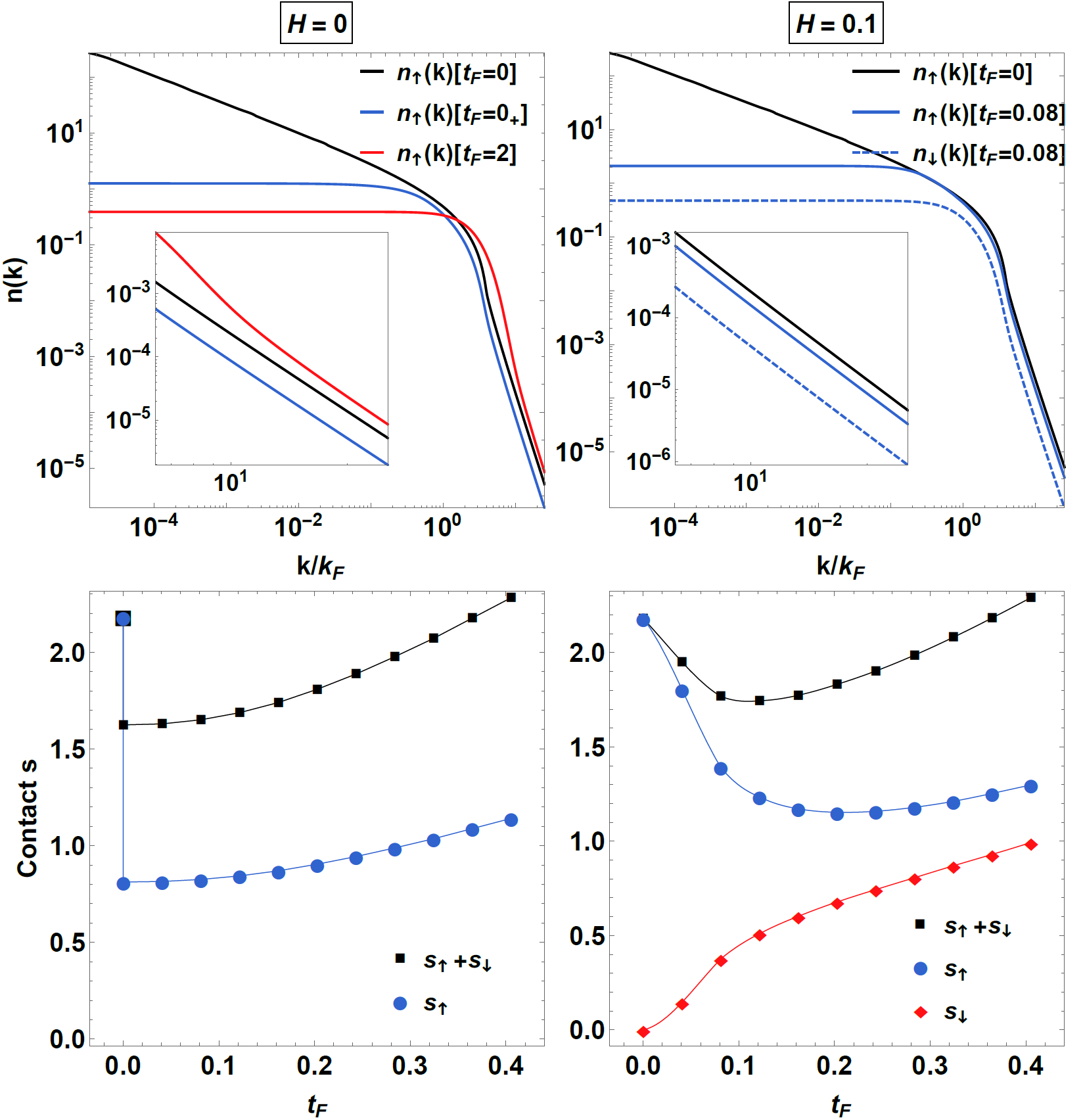}
\caption{ Upper left panel: Momentum distribution in log-log coordinates for
  the balanced impenetrable gas ($H=0$) for $t_F=\{0, 0_+, 2\}$. For the
  balanced system at a fixed temperature $n_\uparrow(k)=n_\downarrow(k)$.
  Data show a significant momentum reconstruction especially at low and high
  momenta as we increase the temperature from the ferromagnetic liquid
  $(t_F=0)$ to the spin-incoherent LL regime $(t_F=0_+)$. The inset shows the
  high momentum tail from which the contact can be extracted and shows that
  $s_\uparrow(t_F=0)>s_\uparrow(t_F=0_+).$ Lower left panel: Dependence of the
  dimensionless total contact $s_\uparrow+s_\downarrow$ (black squares) and
  contact of spin-up particles $s_\uparrow$ (blue disks) as functions of the
  temperature for the balanced system.  While at $t_F=0$ the ground state is
  populated only by type 1 particles, at $t_F=0_+$ the system is balanced and
  $s_\uparrow=s_\downarrow$. The total contact and also $s_\uparrow$ present a
  sharp minimum at $t_F=0_+$.  Upper right panel: Momentum distribution for
  $H=0.1$ and $t_F=\{0, 0.08\}$. In this case for a fixed temperature
  $n_\uparrow(k) \neq n_\downarrow(k)$. Data also show a significant momentum
  reconstruction but not as pronounced as in the balanced case. The inset
  shows that $s_\uparrow(t_F=0)>s_\uparrow (t_F=0.08)>s_\downarrow(t_F=0.08)$.
  Lower right panel: Temperature dependence of the contact for $H=0.1$. Even
  though $s_\downarrow$ is monotonously increasing, the total contact and
  $s_\uparrow$ present a local minimum at low temperatures.
  In all cases $n=1/2$.}
\label{TanMom}
\end{figure}

A universal property of physical systems with contact interactions is that the
large momentum distribution behaves like $\lim_{k\rightarrow\infty}
n_\sigma\sim \mathcal{C}_\sigma/k^4$ with $\mathcal{C}_\sigma$ the amplitude
called contact \cite{Tan1,OD,BZ,CJGZ,PK2}.
In the case of the bosonic Gaudin-Yang model the total contact can be computed
from the thermodynamics \cite{PK2} as $\mathcal{C}=\mathcal{C}_\uparrow+
\mathcal{C}_\downarrow=c^2 \6^2\phi/\6 c^2.$
In Fig.~\ref{TanT} we show the temperature dependence of the dimensionless total
contact $s=\mathcal{C}/k_F^4$ with $k_F=\pi n$ for several values of coupling constant and
magnetic field.
For
$\gamma=\{0.2,1\}$ it is a monotonously increasing function of the
temperature for all values of $H$. At strong coupling the contact develops a
minimum at low temperatures which is more pronounced as $H$
decreases.  This means that for $\gamma\gg 1 $ we encounter the
counterintuitive phenomenon in which the fraction of particles with high
momenta decreases as we increase the temperature. It was first noticed in
\cite{CSZ} for the case of impenetrable particles
at $H=0$ that this momentum reconstruction signals the
transition from the ferromagnetic liquid phase $(T<T_0\sim T_F/ \gamma)$
 to the spin-incoherent LL regime $(T_0<T<T_F)$ \cite{CZ1,FB,F}.
The two temperature scales become well separated for $\gamma\gg 1$ which explains
why we encounter this phenomenon only in the strong coupling limit.

In the impenetrable case the momentum distribution and the contact of each
species of particles can be computed using the Fredholm determinant representation
for the Green's function $\langle\Psi_i^\dagger (x)\Psi_i (0)\rangle_{T,\mu,H}$
obtained in \cite{IP1}.  Its Fourier transform yields the momentum distribution $n_i(k)$ and
the contact is extracted as $C_i=\lim_{k\rightarrow \infty} n_i(k)/k^4.$
The detailed results in Fig.~\ref{TanMom} reveal in addition to the
high momentum crossover also a significant momentum
reconstruction at low-momenta.
This reconstruction, which takes place in a very narrow interval of temperature, is
largest for the balanced system (see the left panels of Fig.~\ref{TanMom}) and
becomes more attenuated as we increase $H$. The low-momentum
crossover has a simple explanation.
At $T=H=0$ the system is described by the impenetrable
Lieb-Liniger model for which the groundstate is a quasicondensate with
$n_\uparrow(k) \sim 1/\sqrt{k}$ for small k (corresponding to the large distance asymptotics
of the Green's function $\sim 1/x^{1/2}$). As we
increase the temperature the system is described by the
spin-incoherent LL with exponentially decreasing Green's function at large
distances \cite{CSZ} and a finite momentum distribution at
low-momenta. The same type of reasoning also holds for strong but finite coupling
implying that the momentum reconstruction both at low- and high-momenta can be
experimentally observed even for moderate values of $\gamma$.

In summary, we have investigated the universal properties of the spinor Bose
gas and identified the universality class and boundaries of the QC region
separating the vacuum from the ferromagnetic liquid phase. The universal contact
increases with increasing magnetic field opposite to the behaviour of fermionic
systems. In the Tonks-Girardeau regime the contact develops a pronounced minimum
indicating a significant momentum reconstruction at low-temperatures which can be
detected in current experiments with ultra-cold gases.

\begin{acknowledgments}
O.I.P.~acknowledges the financial support from the LAPLAS 4 and LAPLAS 5 programs of the
Romanian National Authority for Scientific Research. A.K.~is grateful to DFG
(Deutsche Forschungsgemeinschaft) for financial support in the framework of
the research unit FOR 2316. A.F.~acknowledges CNPq (Conselho Nacional de
Desenvolvimento Cient\'ifico e Tecnol\'ogico) for financial support. All
authors would like to thank the mathematical research institute MATRIX in
Australia where part of this research was performed.
\end{acknowledgments}

\newcommand{\bb}[1]{\boldsymbol{#1}}
\newcommand{\gexp}[1]{\ensuremath{\exp\left\{#1\right\}}}
\newcommand{\mtext}[1]{\qquad \text{#1} \qquad}

\pagebreak
\onecolumngrid
\vspace{\columnsep}
\newpage
\begin{center}
\textbf{\large Supplemental Material: Universality and quantum criticality of the one-dimensional spinor Bose gas}
\end{center}
\vspace{1cm}

\setcounter{equation}{0}
\setcounter{figure}{0}
\setcounter{table}{0}
\renewcommand{\theequation}{S\arabic{equation}}
\renewcommand{\thefigure}{S\arabic{figure}}
\renewcommand{\thetable}{S\arabic{table}}
\renewcommand{\bibnumfmt}[1]{[S#1]}
\renewcommand{\citenumfont}[1]{S#1}
\addtolength{\textfloatsep}{1mm}

\section{Scaling behaviour in the vicinity of the Quantum Critical Point}


We can derive the scaling behaviour of the thermodynamic quantities in the vicinity of the QCP from the
knowledge of the regular part of the pressure $p_r(\mu,H,T)$ and  the universal function $\mathcal{P}_H$.
In the case of the density and compressibility starting from Eq.~(3), introducing the variable
$x=(\mu-\mu_c(H))/ T^{\frac{1}{\nu z}}$ and using $\phi(\mu,H,T)=-p(\mu,H,T)\, ,$  $n=-\6 \phi/\6 \mu$ and
$\kappa=-\6^2\phi/ \6\mu^2$ we find
\be\label{s1}
n(\mu,H,T)\sim \frac{ \6 p_r(\mu,H,T)}{\6 \mu}+T^{\frac{d}{z}+1-\frac{1}{\nu z}}\mathcal{P}'_H\left(x\right)\, ,\ \ \ \
\kappa(\mu,H,T)\sim \frac{\6^2 p_r(\mu,H,T)}{\6\mu^2}+T^{\frac{d}{z}+1-\frac{2}{\nu z}}\mathcal{P}_H^{''}\left(x\right)\, ,
\ee
where $\mathcal{P}'_H(x)$ and $\mathcal{P}_H^{''}(x)$ are the first and second derivative of $\mathcal{P}_H(x)$.
The scaling of the pressure and density ($\tilde p_r=\tilde n_r\sim 0$) which shows that the critical exponents are $z=2$
and $\nu=1/2$ is presented in  Figs.~S1 and S2.
In a similar fashion we can obtain the scaling behaviour for the entropy $s=-\6 \phi/\6 T$  and grand-canonical
specific heat $c_V=-T\6^2 \phi/\6 T^2$
\begin{align}\label{s2}
s(\mu,H,T)&\sim \frac{\6 p_r(\mu,H,T)}{\6 T}+T^{\frac{d}{z}}
\left[\left(\frac{d}{z}+1\right)\mathcal{P}_H(x)-\frac{1}{\nu z}\, x\, \mathcal{P}'_H(x)\right]\, ,\\
c_V(\mu,H,T)&\sim \frac{\6^2 p_r(\mu,H,T)}{\6 T^2}+T^{\frac{d}{z}-1}
\left\{\frac{d}{z}\left(\frac{d}{z}+1\right)\mathcal{P}_H(x)
-\left[\frac{1}{\nu z}\left(\frac{2d}{z}+1\right)-\frac{1}{(\nu z)^2}\right]\, x\, \mathcal{P}'_H(x)
+\frac{1}{(\nu z)^2}\, x^2\,\mathcal{P}^{''}_H(x)
\right\}\, .
\end{align}

From Eqs.~S1 and S3 and taking into account that for the QPT from the vacuum to the ferromagnetic liquid phase
$p_r(\mu,H,T)\sim 0$ we obtain the scaling behaviour of the compressibility Wilson ratio $R_W^\kappa=\frac{\pi^2
k_B^2}{3} T \frac{\kappa}{c_V}\sim \frac{\pi^2 k_B^2}{3}\mathcal{Q}_H(x) $ with
$\mathcal{Q}_H(x)= \mathcal{P}_H^{''}(x)/\left(\frac{3}{4}  \mathcal{P}_H(x)- x \mathcal{P}'_H(x)+x^2
\mathcal{P}_H^{''}(x)\right).$

\vspace{0.5 cm}

\twocolumngrid

\begin{figure}[H]
\includegraphics[width=\linewidth]{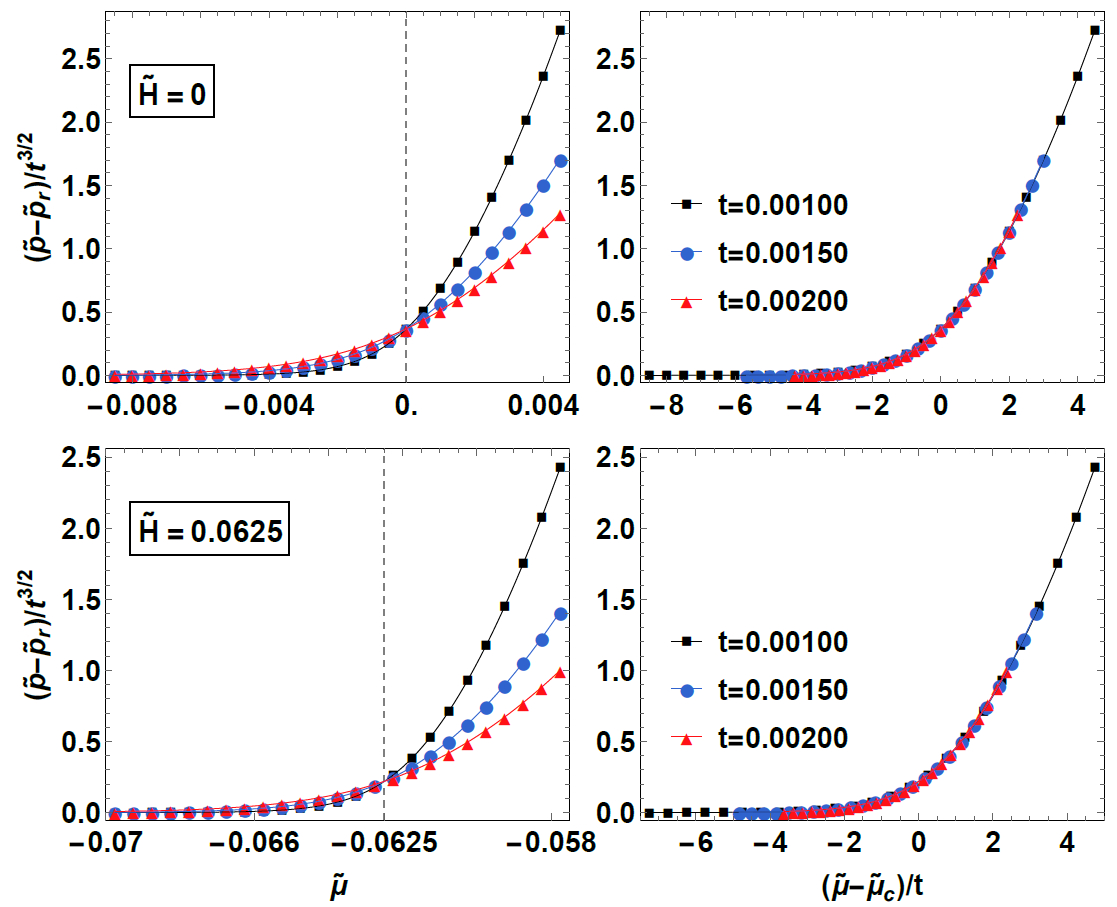}
\caption{Scaling behaviour of the pressure $\tilde p=p/c^3$ for $\tilde{H}=0$ (upper panels) and $\tilde{H}=0.0625$
(lower panels) ($\tilde H=H/c^2, \tilde\mu=\mu/c^2, t=T/c^2$ with $c=2$). The curves at different  temperatures
intersect at the QCP $\tilde{\mu}_c=-|\tilde{H}|$. Plots of the pressure as function of  $(\tilde{\mu}-\tilde{\mu}_c)/t$
reveal the universal function $\mathcal{P}_H(x)$.
}
\label{Pressure}
\end{figure}
\begin{figure}[H]
\includegraphics[width=\linewidth]{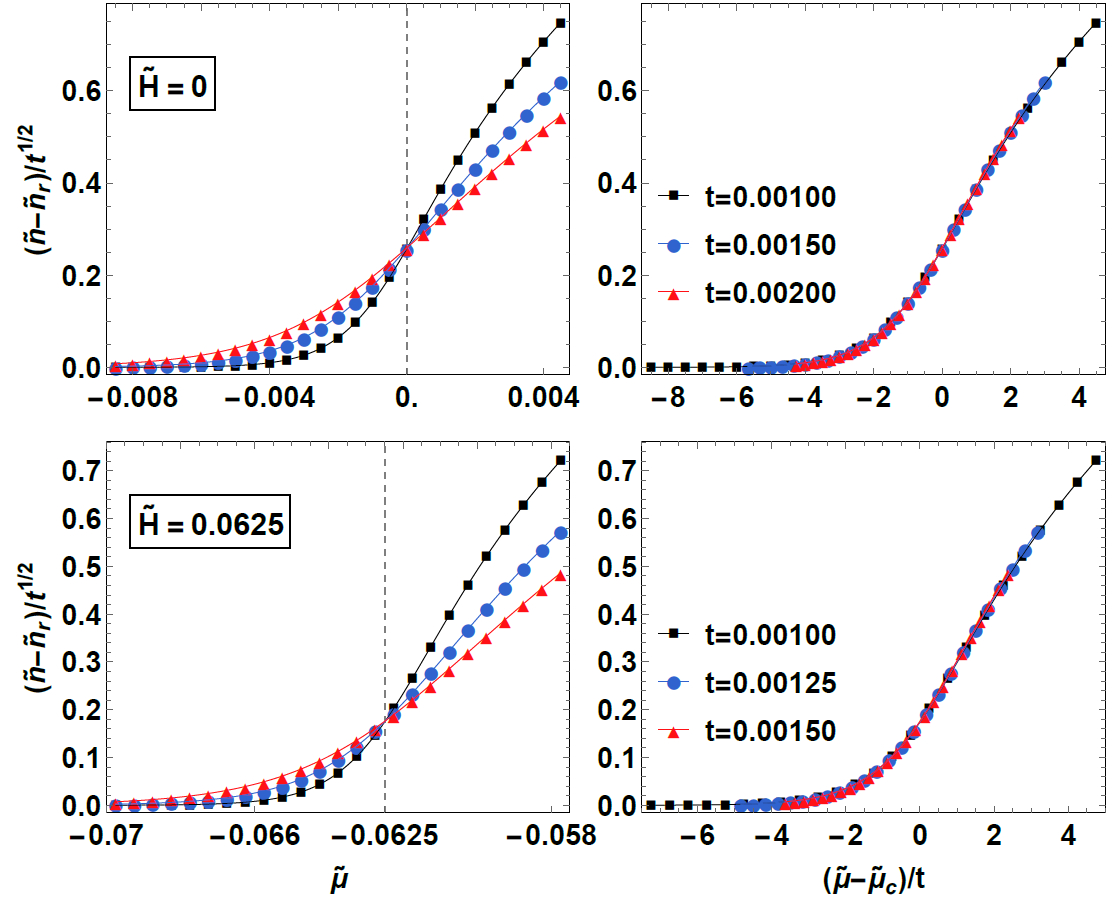}
\caption{Scaling behaviour of the density $\tilde n=n/c$ for $\tilde{H}=0$ (upper panels) and
$\tilde{H}=0.0625$ (lower panels) ($\tilde H=H/c^2, \tilde\mu=\mu/c^2, t=T/c^2$ with $c=2$). The curves at
different  temperatures intersect at the QCP $\tilde{\mu}_c=-|\tilde{H}|$. Plots of the density as function of
$(\tilde{\mu}-\tilde{\mu}_c)/t$   reveal the universal function $\mathcal{P}'_H(x)$.
}
\label{Density}
\end{figure}

\onecolumngrid

\section{Asymptotic analysis of the scaling function}

 Here we investigate the asymptotic properties of  Takahashi's formula
 \be
p=T^{3/2}\mathcal{P}_y(x)\, ,\ \ \ \ \ \ \mathcal{P}_y(x)=\frac{1}{\pi}\int_0^\infty\ln\left[1+(1+e^{-2|y|})
e^{-k^2+x}\right]\, dk\, ,
\ee
which describes the thermodynamics of 1D impenetrable two-component repulsive fermions and can be derived
from the $c\rightarrow\infty$ limit of the TBA equations. While this formula was first obtained in the case
of the impenetrable 2CFG it is also valid for the 2CBG and Bose-Fermi mixture in the same limit.

\textit{The large $x$ limit}. We introduce the notation $\alpha=x+\ln\left(1+e^{-2|y|}\right)$ (large $x$ implies also large $\alpha$)
and successively perform an  integration by parts and changes of variables  $k'= k^2$ and $z=k'-\alpha$ obtaining
\[
\mathcal{P}_y(x)=\frac{1}{\pi}\int_{-\alpha}^\infty\frac{(z+\alpha)^{1/2}}{e^z+1}\, dz\, = \frac{1}{\pi}\int_{-\alpha}^0\frac{(z+\alpha)^{1/2}}{e^z+1}\, dz
+\frac{1}{\pi}\int_0^\infty\frac{(z+\alpha)^{1/2}}{e^z+1}\, dz\,\, .
\]
Switching to $z=-z$ in the first term on the r.h.s. of the previous identity and using $1/(e^{-z}+1)=1-1/(e^z+1)$
we find
\[
\frac{1}{\pi}\int_{-\alpha}^0\frac{(z+\alpha)^{1/2}}{e^z+1}\, dz= \frac{1}{\pi}\int_0^{\alpha}\frac{(\alpha-z)^{1/2}}{e^{-z}+1}\, dz
 = \frac{1}{\pi}\int_0^\alpha(\alpha-z)^{1/2}\, dz-\frac{1}{\pi}\int_0^{\infty}\frac{(\alpha-z)^{1/2}}{e^{z}+1}\, dz+
 \frac{1}{\pi}\int_{\alpha}^{\infty}\frac{(\alpha-z)^{1/2}}{e^{z}+1}\, dz\, ,
\]
where the last term can be neglected because it is exponentially decreasing in $\alpha$.
We have
\begin{align*}
\mathcal{P}_y(x)&=\frac{1}{\pi}\int_0^\alpha(\alpha-z)^{1/2}\, dz+\frac{1}{\pi}\int_0^{\infty}\frac{(\alpha+z)^{1/2}-(\alpha-z)^{1/2}}{e^{z}+1}\, dz\, ,\\
      &=\frac{1}{\pi}\int_0^\alpha(\alpha-z)^{1/2}\, dz+\frac{1}{\pi}\int_0^{\infty}\frac{\alpha^{-1/2} z}{e^{z}+1}\, dz+O(\alpha^{-5/2})\, .\\
      &=\frac{2}{3 \pi}\alpha^{3/2}+\frac{\pi}{12}\alpha^{-1/2}+ O(\alpha^{-5/2})\, ,
\end{align*}
where we have used $\int_{0}^\infty z/(e^z+1)\, dz=\pi^2/12$. Therefore
\begin{align}
\mathcal{P}_y(x)=&\frac{2}{3 \pi}\left[x+\ln\left(1+e^{-2|y|}\right)\right]^{3/2}+\frac{\pi}{12}\left[x+\ln\left(1+e^{-2|y|}\right)\right]^{-1/2}+ O(x^{-5/2})\nonumber\\
=& \frac{2}{3 \pi}x^{3/2}+\frac{\pi}{12} x^{-1/2} \mbox{ for } x\gg 1,\, \  y\gg 1\, ,\label{a2}\\
=&\frac{2}{3 \pi}x^{3/2}+\frac{1}{\pi}x^{1/2}(\ln 2 -|y|)+\frac{1}{4 \pi}x^{-1/2}(\ln^2 2-2\ln 2|y|)+\frac{\pi}{12}x^{-1/2}\, , \mbox{ for } x\gg 1,\, \  y\ll 1\, .\label{a3}
\end{align}
\textit{The small or large negative  $x$ limit}. In this case if $(1+e^{-2|y|})e^{x}$ is smaller than 1  we can use
$\ln(1+x)=x-x^2/2+\cdots$ with the result
\be\label{a4}
\mathcal{P}_y(x)=\pi^{1/2}\sum_{n=1}^\infty \frac{(-1)^{n+1}}{n^{3/2}} \delta^n \, ,\ \ \ \delta=(1+e^{-2|y|})e^x\, ,
\ee
an expansion which is valid for all  $\delta<1$. For large negative $x$ (which characterizes the classical gas (vacuum) regime) only
the first term is necessary.

At low temperatures for $H>0$ the leading term of the magnetization can be computed using Eq.~(\ref{a2}) with the result
\be
m\equiv\lim_{T\rightarrow 0}\frac{\6 p}{\6 H}=\frac{\6}{\6 H}\left[T^{3/2}\frac{2}{3\pi}\left(\frac{\mu+|H|}{T}\right)^{3/2}\right]=\frac{\mbox{sign } H}{\pi}(\mu+|H|)^{1/2}+O(T^2)\, .
\ee
The consequence of the jump in the magnetization is (ground state) ferromagnetism which is consistent with the ferromagnetic property of Bose gases with arbitrary
spin-independent interaction.
In the case of zero magnetic field using Eq.~(\ref{a3}) we obtain for the entropy
\be
s\equiv \lim_{T\rightarrow 0}\frac{\6 p}{\6 T}=\frac{\6}{\6 T}\left[T^{3/2}\frac{\ln 2}{\pi}\left(\frac{\mu}{T}\right)^{1/2}\right]=\frac{\mu^{1/2}\ln 2}{\pi}+O(T)\, .
\ee

\end{document}